\documentclass[prl,aps,twocolumn,floats,showpacs,superscriptaddress,apsrev4-1,letter]{revtex4-1}
\usepackage{graphicx}
\usepackage{amssymb}
\begin{document}
\title{Electronically Guided Self Assembly within Quantum Corrals}

\author{R.~X. Cao}
\affiliation{National Laboratory of Solid State Microstructures and
Department of Physics, Nanjing University, 22 Hankou Rd., Nanjing
210093, China}
\author{B.~F. Miao}
\affiliation{National Laboratory of Solid State Microstructures and
Department of Physics, Nanjing University, 22 Hankou Rd., Nanjing
210093, China}
\author{Z.~F. Zhong}
\affiliation{National Laboratory of Solid State Microstructures and
Department of Physics, Nanjing University, 22 Hankou Rd., Nanjing
210093, China}
\author{L. Sun}
\affiliation{National Laboratory of Solid State Microstructures and
Department of Physics, Nanjing University, 22 Hankou Rd., Nanjing
210093, China}
\author{B. You}
\affiliation{National Laboratory of Solid State Microstructures and
Department of Physics, Nanjing University, 22 Hankou Rd., Nanjing
210093, China}
\author{W. Zhang}
\affiliation{National Laboratory of Solid State Microstructures and
Department of Physics, Nanjing University, 22 Hankou Rd., Nanjing
210093, China}
\author{D. Wu}
\affiliation{National Laboratory of Solid State Microstructures and
Department of Physics, Nanjing University, 22 Hankou Rd., Nanjing
210093, China}
\author{An Hu}
\affiliation{National Laboratory of Solid State Microstructures and
Department of Physics, Nanjing University, 22 Hankou Rd., Nanjing
210093, China}
\author{S. D. Bader}
\affiliation{Materials Science Division, Argonne National
Laboratory, 9700 S. Cass Ave., Argonne, IL 60439}
\author{H.~F. Ding}
\email[Corresponding author: ]{hfding@nju.edu.cn}
\affiliation{National Laboratory of Solid State Microstructures and
Department of Physics, Nanjing University, 22 Hankou Rd., Nanjing
210093, China}

\begin{abstract}
A grand challenge of nanoscience is to master the control of structure and properties in order to go beyond present day functionality. The creation of nanostructures via atom manipulation by means of a scanning probe represents one of the great achievements of the nano era. Here we build on this achievement to self-assemble nanostructures within quantum corrals. The structuring is guided by the quantum confinement of the electronic density of a silver metallic substrate within the corrals. We experimentally demonstrate different self-organized Gd atomic structures confined within 30-nm circular and triangular Fe quantum corrals. This approach enables the creation of model systems to explore and understand new nanomaterials and device prototypes.
\end{abstract}

\maketitle %

Atoms are the building blocks of all materials. The control of the geometric, electronic and magnetic properties of atomic-scale structures provides model systems for the understanding and fabrication of new materials and devices. Presently, two main approaches are used for creating atomic structures. Atomic manipulation enables the building of novel structures in an atom-by-atom fashion~\cite{Eigler-Nature1990,Crommie-Science1993,Monoharan-Nature2000,Hirjibehedin-Science2006,Nilius-Science2002,Loth-Science2012}. It is, however, difficult to build large-scale structures in this manner. Meanwhile, self-organization can provide the formation of atomic-scale structures with relatively large area homogeneity~\cite{Whitesides-Science2002,Chambliss-PRL1991,Brune-Nature1998,Sun-Science2000,Gambardella-Nature2002,Silly-PRL2004,Ding-PRB2007}. Due to its self-limiting formation mechanism, self-organization techniques typically lack the flexibility for local functionality design. As size shrinks to the 10-nm range, quantum effects appear and can influence many system properties. For instance, when electrons are confined in the vertical direction, it has been found that quantum confinement has a decisive influence on the growth of thin films, resulting in novel effects, such as a critical thickness for smooth film growth, and magic heights of nano-islands~\cite{Hinch-EPL1989,Smith-Science1996,Gavioli-PRL1999,Yeh-PRL2000,Luh-Science2001,Oezer-PRB2005}. The underlying mechanism is explained with an electronic growth model~\cite{Zhang-PRL1998}. If quantum confinement in the lateral direction could also be imposed, this would open the possibility for local functionality design with $<$10-nm resolution. Recently, the quantum size effect was proposed to influence self-organization, resulting in new atomic structures, such as a 'quantum onion'~\cite{Stepanyuk-PRL2006}. However, experimental verification of this proposal is still missing.

In this Letter, we experimentally demonstrate that lateral atomic-scale structures can be engineered by exploiting the quantum confinement of surface electronic states provided by the substrate. We utilize scanning tunneling microscopy (STM) manipulation to build circular (diameter: 30 nm) and triangular (linear size: 30 nm) quantum corrals on Ag(111) surfaces with Fe adatoms. The STM spectroscopy measurements of the circular corrals' local density of states (LDOS) around the Fermi energy, $E_F$ show concentric standing waves, as revealed previously~\cite{Crommie-Science1993}. A few Gd atoms are introduced into the corrals and the motions of these atoms are studied. The statistics reveal that the Gd adatom probability distribution inside the corral forms several different trajectories and it is closely related to oscillations of the LDOS at $E_F$. By tuning the Gd coverage, different self-organized Gd atomic structures are formed within circular and triangular Fe quantum corrals. The findings demonstrate that quantum confinement can be used to engineer atomic structures and atom diffusion. Our samples utilize 30-nm diameter circular and triangular corrals fabricated via STM manipulation. But, 30-nm resolution can be reached by means of advanced lithography. Adding quantum engineering to augment it opens new possibilities for local functionality design down to the atomic scale. The method is of advantage in circumventing the difficuties of STM manipulation in building large-area structures and self-organization in functionality design, respectively.

The experiments were performed in an ultrahigh vacuum chamber ($2\times 10^{-11}$~mbar) equipped with a low temperature STM and a sputter gun. The surface of a high purity Ag(111) crystal is cleaned by repeated cycles of argon ion sputtering (at 1.5~keV) and annealing (at 870~K). After that, the crystal is transferred into the STM stage and cooled to 4.7~K. High purity Fe and Gd are deposited by means of electron beam evaporation onto the substrate in the STM stage at $\approx$6~K from outgassed rods. The typical rate of deposition is 0.002~monolayer/min. Electrochemically etched W tips are used for the STM measurements. The bias voltage, $U$, refers to the sample voltage with respect to the tip. Spectroscopy measurements are performed via the modulation technique utilizing a 4-mV amplitude and 6.09-kHz frequency.

\begin{figure}[ht]
\centering
\vskip -1 mm
\includegraphics[width=7.5cm]{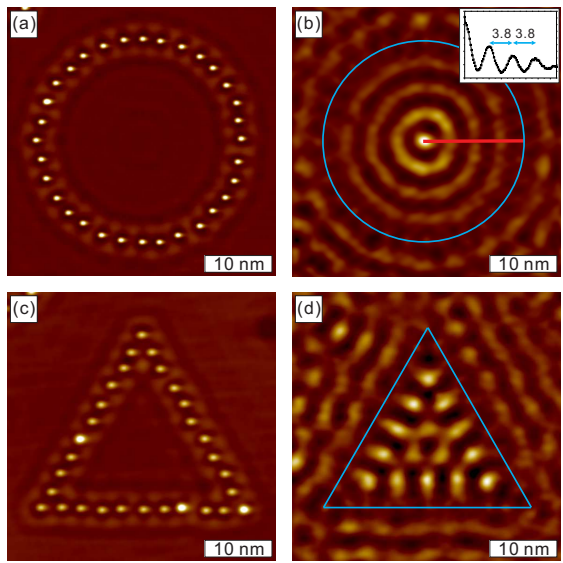}
\caption{(a) STM topography of circular corral built with 32 Fe adatoms on Ag(111) ($U=100$~mV, $I_t=1$~nA). (b) LDOS near $E_F$ inside the corral. Fe circular corral is indicated by blue circle for clarity. Inset: average period of standing wave indicated by red line in (b) is $\approx 3.8$~nm which corresponds to half of the Fermi wavelength of Ag(111). (c) and (d) are topography ($U=100$~mV, $I_t=1$~nA) and LDOS near $E_F$ of a triangular corral, respectively.} \label{Fig_1}
\vskip -5 mm
\end{figure}

We choose Fe adatoms as the building blocks of the quantum corrals because of the favorable diffusion barrier of $\approx 43$~meV for Fe on Ag(111)~\cite{Zhang-PRB2010}. With this magnitude barrier, atomic manipulation can be readily achieved, while the adatoms can then be immobilized after positioning at 4.7 K. No apparent change of the corral shape is found for temperature $<14$~K. About 0.001 monolayer equivalents (MLE) of single Fe adatoms are deposited. With the tip stabilizing condition of  1~V and 1~nA, the tip further approaches the sample surface within a distance of $\approx 0.4$~nm to drive the Fe adatoms to the designed positions. Figure~\ref{Fig_1}(a) shows a typical circular quantum corral with 30-nm diameter. The Ag(111) surface contains a two-dimensional (2D) electron gas. These electrons inhabit a surface-state band that starts at 67~meV below $E_F$~\cite{Li-PRB1997}. Surface-state electrons are confined within the corral by strong scattering at the corral walls. The quantum interference between electronic waves traveling toward corral walls and those that are backscattered leads to a concentric circular standing-wave pattern. This kind of pattern can be observed by the spectroscopy image as shown in Fig.~\ref{Fig_1}(b) for the energy near $E_F$. An average line profile from the center to the Fe circle, indicated by the red line, is plotted in the inset of Fig.~\ref{Fig_1}(b). The average period of the standing wave is $\approx 3.8$~nm, which corresponds to half of the Fermi wavelength of Ag(111) surface state. To study the effect of quantum confinement within different shapes, we also build triangular quantum corrals with linear dimension of 30 nm, as presented in Fig.~\ref{Fig_1}(c). The corresponding spectroscopy near $E_F$ in Fig.~\ref{Fig_1}(d) shows a discrete triangular pattern instead of concentric circles. The different patterns inside the different corrals suggest the possibility to design, control and engineer atomic structures via lateral quantum confinement.

\begin{figure}[ht]
\centering
\includegraphics[width=7.65cm]{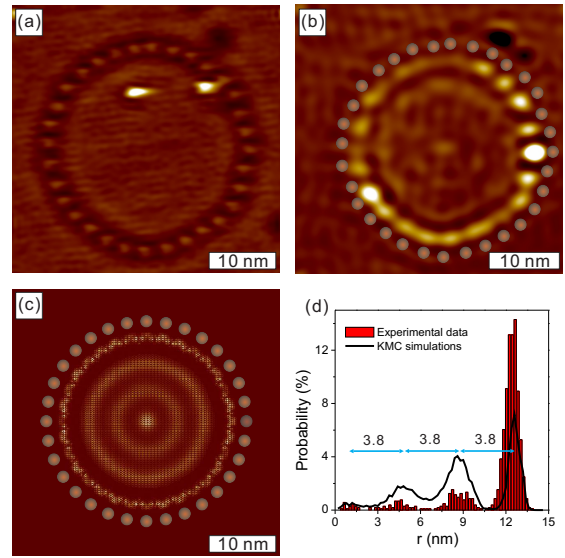}
\caption{(a) Typical STM image of two Gd adatoms inside a circular corral. (b) Adatom probability distribution derived from summing up 510 successive images as in (a). For clarity, small balls are added to mark the positions of Fe adatoms. (c) Adatom probability distribution of KMC simulations with $10^8$ samplings. (d) Experimental histogram (red column) of the radial distributed probability, and theoretical radial distribution (black line) obtained from KMC simulations. Note that the average period of 3.8~nm is observed in both distributions.} \label{Fig_2}
\vskip 2 mm
\end{figure}

Gd is chosen to study the adatom diffusion inside the quantum corrals because Gd has a large magnetic moment and we experimentally find that Gd adatoms can form a well-ordered superlattice on Ag(111) surface, similar to the Ce/Ag(111) system~\cite{Silly-PRL2004}. The superlattice is formed by the long range interaction (LRI) through the surface state electrons~\cite{Silly-PRL2004,Lau-SurfSci1978,Hyldgaard-JPCM2000,Knorr-PRB2002}. About 0.02\%~MLE of Gd adatoms are deposited on the pre-patterned Ag(111) surface and adatom diffusion on the flat terrace and inside the corrals are studied. Temperature dependent measurements of single adatom diffusion show that Gd adatoms have a diffusion barrier of $\approx 7.6$~meV and attempt frequency of $\approx 2\times 10^9$~Hz on Ag(111) and they are mobile and difficult to observe at 4.7~K. Hence, we further cool the system to $<4.4$~K for the diffusion study. To minimize tip-induced atomic motion, we choose tunneling conditions of  50~mV and 2~pA for imaging. On the flat terrace without quantum corrals, we find the Gd adatoms follow a 2D random walk, similar to that observed for the diffusion of Cu single adatoms on a flat Cu(111) surface~\cite{Negulyaev-PRL2008}. For the single Gd adatoms inside the Fe corral, we find that the adatoms mostly stay in the vicinity of a specific location and form an arc-shaped distribution near one side of the Fe corral (Supplementary Movie S1), which is in contrast to the theoretically predicted circular orbits~\cite{Stepanyuk-PRL2006}. This may be due to the fact that the Fe adatoms in the experiments are not positioned in as perfect a circle as the theory assumed, resulting in preferred occupation sites. To identify this, we repeat the same measurements with newly constructed corrals and find that the arcs are always located near one side of the corral, but at different locations for different experiments. The random distribution of these arcs suggests they are preferred occupation sites.

To overcome the problem of Gd adatom trapping at preferred occupation sites, we studied the diffusion of pairs of Gd adatoms inside a quantum corral. The idea is to use the collision of two adatoms to kick an adatom out of the preferred occupation site when it is trapped. Figure~\ref{Fig_2}(a) presents a typical image of two adatoms inside a circular corral. The Gd atoms appear to be larger than the Fe atoms, which is partly attributable to Gd movement during the imaging. To obtain statistics on the Gd diffusion inside the corral, we continue to image the same area until the liquid He in the cryostat evaporates. We collected 510 images (Supplementary Movie S2) and averaged them into a single image. With further background subtraction to remove the electronic effect, the statistical result is shown in Fig.~\ref{Fig_2}(b). The averaged image shows three concentric orbits of adatom motion and one focused at the center. The orbits have different intensities, suggesting that they have different occupancies. The brighter the orbit is, the higher the occupancy is. On the outermost orbit of Fig.~\ref{Fig_2}(b), there are a few bright spots with much higher probability. The appearance of the bright spots relates to the observation of the preferred occupation sites for the single adatom diffusion mentioned above.

\begin{figure}[ht]
\centering
\vskip -13mm
\includegraphics[width=9cm]{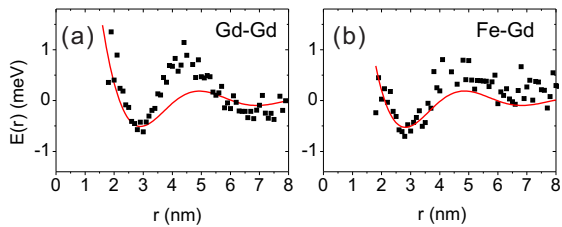}
\vskip -23mm
\caption{Long range interaction between two Gd atoms (a), and Gd-Fe single adatoms (b) on Ag(111). Symbols are experimental data and lines are the fit.} \label{Fig_3}
\end{figure}

In order to compare with the experiments, we performed kinetic Monte Carlo (KMC) calculations to simulate the diffusion process of one and two Gd adatoms inside the corral. The method had been used previously for the simulation of the Fe superlattice formation on Cu(111)~\cite{Zhang-PRB2010}. In the simulations, the hopping rate of an adatom from site \emph{i} to site \emph{j} on the Ag(111) surface is calculated using the expression $\upsilon_{i\rightarrow j}=\upsilon_{0}\exp {(-E_{i\rightarrow j}/k_BT)}$, where $T$ is the temperature of the substrate, $\upsilon_{0}$ is the attempt frequency, $k_B$ is the Boltzmann constant, and $E_{i\rightarrow j}$ is the hopping barrier. The influence of the long-range interaction (LRI) through the surface state electrons is included in the hopping barrier,\emph{ i.e.}, $E_{i\rightarrow j}=E_D+0.5(E_i-E_j)$, where $E_D$ is the diffusion barrier for an isolated atom on a clean surface, and  $E_i(E_j)$ is the total energy caused by the LRI. The LRI between Gd-Gd and Fe-Gd are obtained from a combination of experimentally measured results (Fig.~\ref{Fig_3}) and the fitted curve using the theoretical model of interaction mediated by a Shockley surface-state band~\cite{Hyldgaard-JPCM2000}. The simulated results for two Gd adatoms' probability distribution are shown in Fig.~\ref{Fig_2}(c). It shows three concentric orbits and one focused center, in agreement with the experimental findings. We note that we did not find any apparent difference between the probability distribution for single Gd-adatom or two-adatom diffusion when the Fe adtoms are positioned almost as a perfect circle. To make a more quantitative comparison, we plot both the experimental (red column) and simulated (black curve) radial distribution of the visiting probability in Fig.~\ref{Fig_2}(d). Both data sets show four peaks with almost the same peak positions. The separations between the peaks are all $\approx 3.8$~nm. The difference in the intensities of these peaks may be due to the fact that the Fe adatoms are not ideally positioned as they are in the simulations. In addition, the experimental data are obtained from only 510 images, while the simulated data is the statistical result of $10^8$ samplings. In comparison with the LDOS near $E_F$ shown in Fig.~\ref{Fig_1}(b), one notices the similarity between the LDOS and the distribution of the adatom diffusion, which both show three concentric orbits and one focused at the center. The separations between the orbits are all $\approx 3.8$~nm, which corresponds to half of the Fermi wavelength of the Ag(111) surface state. A difference in the orbit positions is expected as the Gd adatoms will add an additional phase shift. Nevertheless, the similarity between the distribution of the adatom diffusion probability and the LDOS demonstrates that the quantum confinement significantly modify the atomic diffusion.

\begin{figure}[ht]
\centering
\includegraphics[width=8.5cm]{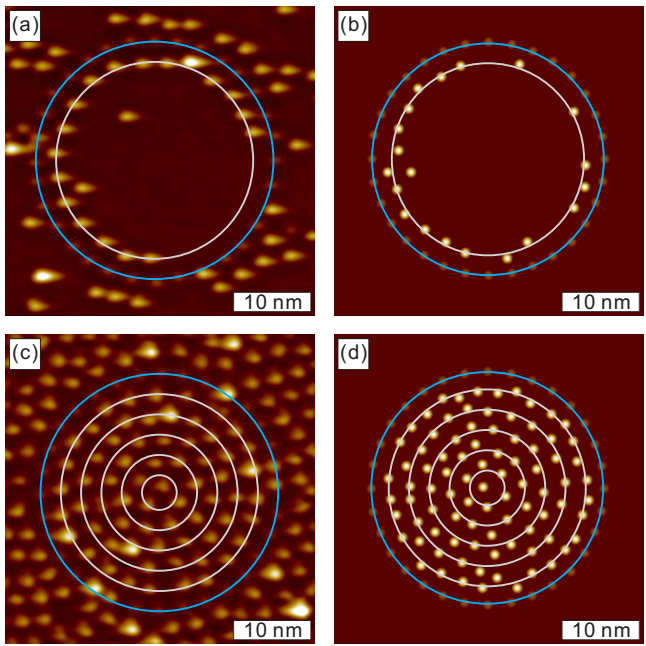}
\caption{(a) Typical topography image of 20 Gd atoms inside an Fe circular corral. (b) KMC simulation of the same coverage. (c) Typical topography image of Gd adatoms with optimal coverage inside a Fe corral. (d) KMC simulation of Gd adatoms with optimal coverage. For clarity, blue and white circles indicate the Fe corral and orbits of Gd adatoms, respectively.} \label{Fig_4}
\end{figure}

In the following, we further demonstrate the possibility to use quantum confinement to create novel atomic structures. For this, we increased the Gd dosage. Figure~\ref{Fig_4}(a) shows a typical image of 20 Gd adatoms inside the Fe corral. We find that most of the adatoms are located near the quantum corral, forming a ring-like structure. This is in agreement with the above diffusion study which shows that the orbit nearest to the Fe corral has the maximum occupancy for the Gd adatom's visit. Figure~\ref{Fig_4}(c) presents a typical image obtained with the optimal dosage. In sharp contrast with the superlattice on flat terraces, we find that the Gd adatoms form five concentric orbits inside the corral, as marked by the white concentric circles. Except for a few brighter spots, most (60 out of 66 single adatoms) of the Gd adatoms are located within the orbits. The brighter spots are Gd dimers or trimers that formed during the deposition. The formation mechanism of dimers and trimers was discussed previously~\cite{Zhang-PRB2010}. The dimer/trimer typically has a much higher diffusion barrier than a single Gd adatom. Therefore, they are anchored once they form, and their positions may not be inside the orbits. The orbits have almost uniform separation of about 2.5~nm, in contrast to 3.8~nm found in the diffusion study mentioned above. This can be understood from the following. The Gd adatoms initially occupy the orbit nearest to the Fe quantum corral and this orbit has a separation of 2.5~nm from the corral. After filling this orbit, the newly added Gd adatoms feel the potential created by the new and smaller Gd corral instead the Fe corral. From the LRI measurements (Fig.~\ref{Fig_3}), the interaction energy between two Gd adatoms has an energy minimum at $\approx 2.9$~nm. Therefore, the Gd adatoms prefer to have a separation of 2.9~nm to each other within the orbit. The lowest energy minimum position created by these two Gd adatoms forms an equilateral triangle with them. A newly added nearby Gd adatom will take this energy minimum position and have a separation of $2.9\times \sqrt{3}/2 \approx 2.5$~nm to the orbit where these two Gd adatoms locate. Following this simple rule, every newly formed orbit will have a radius of $\approx 2.5$~nm smaller than the previous one. We also performed KMC simulations with corresponding Gd coverage according to the experiments. The results are shown in Figs.~\ref{Fig_4}(b) and (d) and they are in agreement with the experiments.

\begin{figure}[ht]
\centering
\includegraphics[width=7cm]{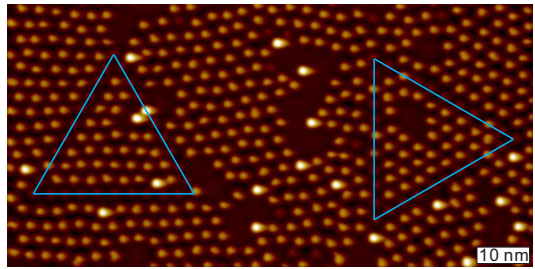}
\caption{Self-organization of Gd adatoms inside triangular corrals with different directions. For clarity, Fe corrals are marked by blue triangles.} \label{Fig_5}
\end{figure}

Quantum engineering requires to tailor the structures with arbitrary shapes. Above, we demonstrate the capability with circular quantum confinement. The question remains whether we can use this method to create patterns with other shapes. To test this, we further constructed two equilateral triangles with 60 Fe adatoms. The lateral size of the corrals is 30 nm. The corrals are separated from each other by $\approx 55$~nm. One triangular corral is intentionally rotated by $30^\circ$ with respect to the other. After deposition of Gd with carefully tuned coverage, we find that the Gd adatoms form well-ordered patterns in both corrals (Fig.~\ref{Fig_5}). The Gd adatoms are aligned parallel to the triangle side. In addition, the structures created inside the triangular corral follow the orientation of the corral. This demonstrates that the orientation of the novel structures can also be controlled. The findings are also supported by the KMC simulations (not shown). The patterns are clearly different from the concentric circular pattern formed inside the circular corrals discussed above. This again demonstrates the use of quantum confinement to create novel atomic structures.

In summary, we experimentally demonstrate that one can use quantum confinement to control atom positioning and create novel atomic structures. Our examples utilize 30-nm diameter circular and triangular corrals fabricated via STM manipulation. But, 30-nm resolution can be reached by means of advanced lithography. The message is that it can be utilized in conjunction with quantum engineering to open new possibilities for the local functionality design down to the atomic scale. We note that this capability is not limited on this particular system only, \emph{e.g.}, Ce adatoms on Ce-dimmer patterned Ag(111) was also proposed~\cite{Stepanyuk-PRL2006}.

\acknowledgments Work at Nanjing is supported by the State Key Program for Basic Research of China (Grant No. 2010CB923401), NSFC (Grants Nos. 10974087, 10834001, and 11023002) and PAPD. Work at Argonne is supported by the U.S. Department of Energy, Office of Science, Basic Energy Sciences, under contract No. DE-AC02-06CH11357.

\end{document}